\documentclass[openacc]{rstransa}

\begin{document}

\title{Particle physics experiments based on the AWAKE acceleration scheme}

\author{
M. Wing$^{1}$}

\address{$^{1}$Department of Physics and Astronomy, UCL, Gower Street, London WC1E 6BT, UK}

\subject{Particle physics, high-energy physics, plasma physics}

\keywords{proton-driven plasma wakefield acceleration, dark photons, strong-field QED, electron--proton scattering}

\corres{M. Wing\\
\email{m.wing@ucl.ac.uk}}

\begin{abstract}
New particle acceleration schemes open up exciting opportunities, potentially providing more 
compact or higher-energy accelerators.  The AWAKE experiment at CERN is currently taking 
data to establish the method of proton-driven plasma wakefield acceleration.  A second phase  
aims to demonstrate that bunches of about $10^9$\,electrons can be accelerated to high energy, 
preserving emittance and that the process is scalable with length.  With this, an electron beam of 
${\mathcal O}$(50\,GeV) could be available for new fixed-target or beam-dump experiments 
searching  for the hidden sector, like dark photons.  The rate of electrons on target could be 
increased by a factor of more than 1000 compared to currently available, leading to a 
corresponding increase in sensitivity to new physics.  Such a beam 
could also be brought into collision with a high-power laser and thereby probe the completely 
unmeasured region of strong fields at values of the Schwinger critical field.   An ultimate goal 
is to produce an electron beam of ${\mathcal O}$(3\,TeV) and collide with an LHC proton beam. This very high 
energy electron--proton collider would probe a new regime in which the structure of matter is 
completely unknown. 
\end{abstract}


\begin{fmtext}
\section{Introduction}

The Standard Model of particle physics~\cite{np:22:579, prl:19:1264,salam-nobel,np:b44:189} is amazingly successful 
in describing the fundamental particles of nature and their interactions.  It has been built up over 50\,years of experimentation 
and theoretical development with the prediction~\cite{prl:13:321,pl:12:132,prl:13:508,prl:13:585,pr:145:1156,pr:155:1554} and discovery~\cite{pl:b716:1,pl:b716:30} of the Higgs boson being the latest compelling success.  The Standard Model provides 
predictions and can describe a wide range of phenomena, some to tremendous  precision and/or covering a huge kinematic range.

\end{fmtext}


\maketitle

However, there is still much unexplained for which new experiments are needed.  Examples of some of the big questions in 
particle physics are: Why is there so much matter compared to anti-matter in the Universe\,?  What is the nature of dark matter 
(and dark energy)\,?  Can we unify the forces\,?  What is the fundamental structure of matter\,?  These and other  
areas of particle physics have often relied on the use of high-energy particle beams and colliders to resolve and push back the 
boundaries of our knowledge.  The discovery of the Higgs boson, mentioned in the previous paragraph, is one of the latest 
examples of how particle colliders have been key in elucidating the biggest questions in particle physics.
 
In the 1960s, electron--positron colliders had a centre-of-mass energy of about 1 GeV and in the 30\,years up to LEP\,II, in the 1990s, 
the centre-of-mass energy increased up to about 200\,GeV.  A similar increase was also obtained for hadron colliders, culminating 
in the Large Hadron Collider (LHC)~\cite{lhc} which collided protons at a centre-of-mass energy of 13\,TeV for the first time in 2015.  Future 
electron--positron colliders up to and beyond the TeV scale, such as at the International Linear Collider (ILC)~\cite{ilc} or Compact Linear Collider 
(CLIC)~\cite{clic}, will extend to $30-50$\,km.   The use of RF cavities in conventional accelerators, 
which are limited to accelerating gradients of at most 100\,MV/m, means that increased energy such as the ILC or CLIC requires increased 
lengths.  In order to reduce the length of future colliders, novel acceleration techniques are required in which significantly higher accelerating 
gradients are obtained.

Plasma wakefield acceleration~\cite{Tajima:1979bn,Chen:1984up,pop-14-055501,RevModPhys.81.1229,doi:10.1142/S1793626816300036} 
is one such technique that could lead to shorter or higher-energy accelerators.  Pioneering experiments have shown that an intense laser pulse~\cite{Modena:1995dgw,Mangles:2004ta,Geddes:2004tb,Faure:2004tc} or electron bunch~\cite{Blumenfeld:2007ph,Litos:2014} traversing a plasma, drives electric fields of 10s GV/m and above.  Proton-driven plasma wakefield acceleration is well suited to high-energy physics 
applications given the high stored energies possible in proton bunches and hence the possibility to accelerate electrons to high energy in 
one stage~\cite{Caldwell:2008ak}.  However, plasma wakefield acceleration has a number of challenges to overcome: a high repetition 
rate and a high number of particles per bunch is needed so as to maximise luminosity; likewise, the bunch spatial extent needs to be small 
(nm scale) to maximise the luminosity; and the production of beams needs to be efficient and highly reproducible.  The ultimate goal of 
plasma wakefield acceleration is then to have high-quality electron beams at the TeV scale produced over the km length scale.

The advanced wakefield (AWAKE) experiment~\cite{ppcf:56:084013,nim:a829:3,nim:a829:76,ppcf:60:014046} has already demonstrated 
that protons can drive plasma wakes~\cite{arxiv:1809.01191,arxiv:1809.04478} and that electrons can be accelerated to high 
energies~\cite{nature:2018} (see Section~\ref{sec:awake} for more details on the AWAKE experiment).  Given this, it is appropriate to 
think of first, realistic applications of the AWAKE scheme of proton-driven plasma wakefield acceleration.  In having an ultimate goal of 
generating high-quality TeV electron beams, this document considers applications of beams of lower energy and less stringent demands on 
the quality, whilst making strong use of the current CERN infrastructure.  In doing this, the technology will be tested and should enable the 
ultimate aims to be achieved faster, whilst doing experiments that have a novel and exciting particle physics programme.

This paper is organised as follows.  The AWAKE experiment, its results and plans are briefly given in Section~\ref{sec:awake} to outline 
how the scheme will mature towards a useable technology.  The main body of the paper, Section~\ref{sec:expts}, outlines possible 
particle physics experiments that could be realised using the AWAKE scheme, some on timescales of within the next 10\,years.  Some 
possibilities are just stated, with the more promising possibilities discussed in more detail.  The ideas are then briefly summarised in 
Section~\ref{sec:summary}.

\section{An AWAKE-like beam for particle physics experiments}
\label{sec:awake}

The AWAKE experiment is a proof-of-principle project, approved by CERN in 2013, to demonstrate proton-driven plasma wakefield 
acceleration for the first time.  A special extraction of the Super Proton Synchrotron (SPS) provides bunches of 400\,GeV protons, containing a charge of about 
$3 \times 10^{11}$\,protons, up to every 7\,sec to the experimental area.  Experimentation started in 2016 and continued in 2017 in 
which the proton bunch was injected into a rubidium vapour plasma source (with density, 
$n_{pe} = (1 - 10) \times 10^{14}$\,cm$^{-3}$).  Given the length of the SPS proton bunches, $\mathcal{O}(10\,{\rm cm})$, the 
experiment relies on the self-modulation effect which splits the long proton bunch into shorter higher density micro-bunches, 
spaced by the plasma wavelength.  The micro-bunching of the protons was seen in the first data-taking period and the effect has recently 
been published~\cite{arxiv:1809.04478}.  The effect of self-modulation also leads to the transverse expulsion of protons, which was 
also measured in the first data-taking period as an expanded beam halo downstream of the plasma cell and also recently 
published~\cite{arxiv:1809.01191}.  At the end of 2017 and during 2018, experiments were performed in which an external bunch 
of electrons were injected in the wake of the proton beam and accelerated.  At low plasma densities, clear and consistent signals of 
accelerated electrons were seen at energies around $500-800$\,MeV; at higher densities, the number of electrons accelerated was 
smaller, but acceleration up to 2\,GeV was observed~\cite{nature:2018}.  These results demonstrate proton-driven plasma wakefield 
acceleration for the first time and motivate further experimentation to develop the AWAKE scheme as a useable acceleration 
technique.

The AWAKE Run 1 programme finishes in November 2018, at which point the CERN accelerator complex shuts down for two years 
for upgrade and maintenance.  The SPS will start up again in 2021 and run for four years until 2024 and an ambitious AWAKE Run 2 
programme is being developed for this period.  The final goal by the end of AWAKE Run 2 is to be in a position to use the AWAKE 
scheme for particle physics experiments.  In order to achieve this, the aims of AWAKE Run 2 are to have high-charge bunches of 
electrons accelerated to high energy, about 10\,GeV, maintaining beam quality through the plasma and showing that the process is 
scalable.  The parameters are summarised in Table~\ref{table:run2}~\cite{adli:ipac16}.  This will require development of the 
initial electron source, beam and plasma diagnostics as well as development of the plasma technology which can fulfil these 
ambitious goals.   

\begin{table}[!h]
\begin{center}
\caption{Preliminary AWAKE Run 2 electron beam parameters.}
\label{table:run2}
\begin{tabular}{ll}
\hline
Parameter & Value \\
\hline
Accelerating gradient & $> 0.5$\,GV/m \\
Energy gain & 10\,GeV \\
Injection energy & $\gtrsim 50$\,MeV \\
Bunch length (rms) & $40 - 60$\,$\mu$m \\
Peak current & $200 - 400$\,A \\
Bunch charge & $67 - 200$\,pC \\
Final energy spread & few \% \\
Final emittance & $\lesssim10$\,$\mu$m \\\hline
\end{tabular}
\vspace*{-4pt}
\end{center}
\end{table}

\section{Possible particle physics experiments}
\label{sec:expts}

A high-energy electron beam with high-charge bunches from tens of GeV up to TeV energies has many potential 
applications.  Three are outlined in detail in the following sub-sections, but other possibilities are briefly discussed 
here.  A condition of any application is that the particle physics goals must be new, interesting and do something 
not done elsewhere.  An oft mooted application of plasma wakefield acceleration is the development of a high energy, 
high luminosity linear $e^+e^-$ collider.  However, such a collider is a challenge for conventional accelerators and 
so to have this as the first application of plasma wakefield acceleration is ambitious.  Hence, the approach taken here 
is to consider experiments, such as fixed-target experiments and an electron--proton collider, which have less 
stringent requirements on the quality of the beam.  A natural progression would be to build such accelerators before 
attempting to develop a high energy, high luminosity linear $e^+e^-$ collider.  In this way, the accelerator technology 
can be developed whilst still carrying out cutting-edge particle physics.

A high-energy electron beam could be used as a test-beam facility for either detector or accelerator studies.  There 
are not many such facilities world-wide and they are often over-subscribed.  The characteristics of the electron 
beam which would make it rather distinct is the high energy which can be varied; a pure electron beam with low 
hadronic backgrounds; a high bunch charge; and longitudinally short bunches.  These properties may not be ideal 
or will be challenging for detector studies which usually rely on single particles.  However, as an accelerator test 
facility the bunched structure and flexibility in energy will be useful properties.

A natural avenue of study with a high-energy electron beam is the deep inelastic scattering of electrons off protons 
or ions in order to study the fundamental structure of matter~\cite{dis:book}.  The simplest experimental configuration 
is where a lepton beam impinges on a fixed target and many such experiments have been performed in the past.  So 
far only one lepton--hadron collider, HERA~\cite{hera}, has been built.  Potential physics that could be studied at a future deep inelastic 
scattering experiment are to measure the structure of the proton at high momentum fraction of the struck parton in 
the proton, which could be valuable for the LHC, and to understand the spin structure of the nucleon, which is still poorly 
known~\cite{mpl:a24:1087}.  A thorough survey of previous as well as planned experiments 
must be carried out to assess the potential of a deep inelastic scattering fixed-target experiment based on a high-energy electron 
beam (${\mathcal O}(50)$\,GeV) from AWAKE.

\subsection{Experiment to measure strong-field QED at the Schwinger critical field}

The theory of electromagnetic interactions, quantum electrodynamics (QED), has been studied and tested in numerous 
reactions, over a wide kinematic range and often to tremendous precision.  The collision of a high-energy electron 
bunch with a high-power laser pulse creates a situation where QED is poorly tested, namely in the strong-field regime.  In 
the regime around the Schwinger critical field, $\sim 1.3 \times 10^{18}$\,V/m, QED becomes non-linear and these values 
have so far never been achieved in controlled experiments in the laboratory.  Investigation of this regime could lead to a 
better understanding of where strong fields occur naturally such as on the surface of neutron stars, at a black hole's event 
horizon or in atomic physics.

\begin{figure}[!h]
\centering\includegraphics[width=8cm,trim={7cm 12cm 7cm 0cm},clip]{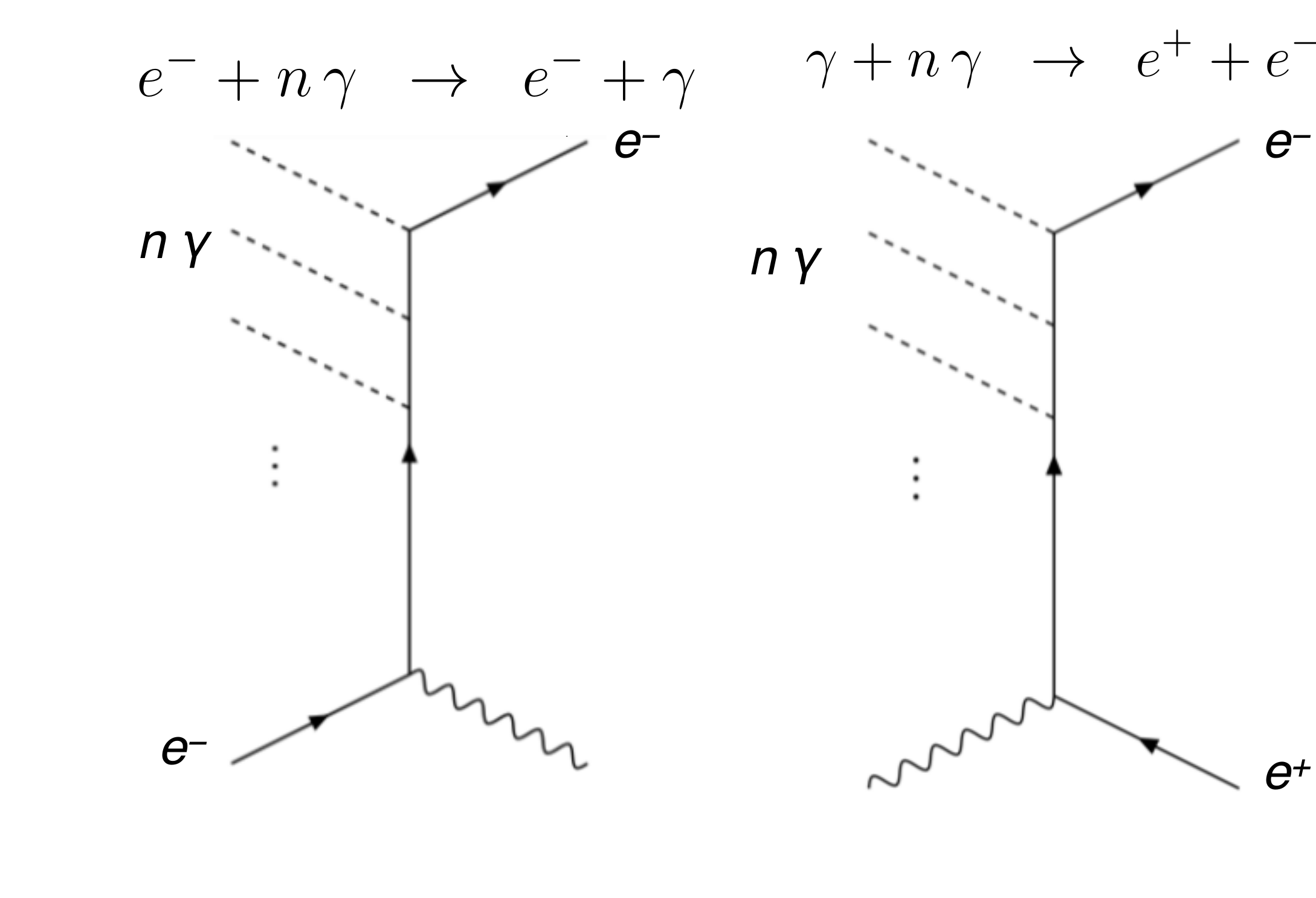}
\caption{A Feynman diagram representation of (a) Compton scattering of an electron and (b) production of an 
$e^+e^-$ pair in the field of a high-power laser in which absorption of multiple photons has taken place.}
\label{fig_qed}
\end{figure}

In the presence of strong fields, rather than the simple $2 \to 2$ particle scattering, e.g.\ $e^- + \gamma \to e^- + \gamma$,  
multi-particle absorption in the initial state is possible, e.g.\ $e^- + n\gamma \to e^- +\gamma$, where 
$n$ is an integer (see Fig.~\ref{fig_qed}).  Therefore an electron interacts with multiple photons in the laser pulse and a photon can also 
interact with multiple photons in the laser pulse to produce an $e^+e^-$ pair, also shown in Fig.~\ref{fig_qed}.  For more details 
on the processes and physics, see a recent review~\cite{ijmp:a33:1830011}.

The E144 experiment~\cite{pr:d60:092004} at SLAC investigated electron--laser collisions in the 1990s using bunches of 
electrons, each of energy about 50 GeV, but due to the limitations of the laser, they did not reach the Schwinger critical 
field.  With the advances in laser technology over the last 20\,years, these strong fields are now in reach~\cite{sqed-ws}.  
However, the current highest-energy electrons are delivered by the European XFEL at 17.5\,GeV and the AWAKE scheme 
has the possibility to provide a higher-energy electron beam which would then be more sensitive to the $e^+e^-$ pair production 
process and probe a different kinematic regime.  

\subsection{Experiment to search for the dark sector}

Dark photons~\cite{JETP:56:502,pl:b136:279,pl:b166:196} are postulated particles which could provide the link to a dark or hidden sector 
of particles.  This hidden sector could explain a number of issues in particle physics, not least of which is that they are candidates 
for dark matter which is expected to make up about 80\% of known matter in the Universe.  Dark photons are expected to have low masses 
(sub-GeV) and couple only weakly to the Standard Model particles and so would have not been seen in previous experiments.  
The dark photon, labelled $A^\prime$, is a light vector boson which results from a spontaneously broken new gauge symmetry 
and kinetically mixes with the photon and couples to the electromagnetic current with strength $\epsilon \ll 1$.  Recently, experimental 
and theoretical interest in the hidden sector has increased and is discussed in recent reviews on the 
subject~\cite{arnps:60:405,us-cosmic-visions}.  

\begin{figure}[!h]
\centering\includegraphics[width=8cm]{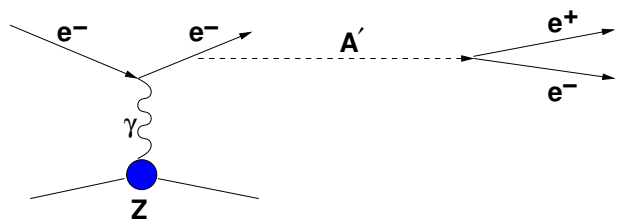}
\caption{A representation of the production of a dark photon, $A^\prime$, in a fixed-target experiment with an electron beam.  
The dark photon subsequently decays to an $e^+e^-$ pair.}
\label{fig_darkphotons}
\end{figure}

A common approach to search for dark photons is through the interaction of an electron with a target in which the dark photon is 
produced and subsequently decays.  This process is shown in Fig.~\ref{fig_darkphotons} in which the dark photon decays to an 
$e^+e^-$ pair.  The NA64 experiment is already searching for dark photons using high-energy electrons on a 
target~\cite{pr:d89:075008,arxiv:1312.3309,prl:118:011802}, initially measuring the dark photon decaying to dark matter particles 
(``invisible mode'')
and so leaving a signature of missing energy in the detector.  Although high-energy electrons of 100\,GeV are used, a limitation of 
the experiment is the rate of electrons is below about $10^6$\,electrons per second as they are produced in secondary interactions 
of the SPS proton beam.

\begin{figure}[!h]
\centering\includegraphics[width=8cm]{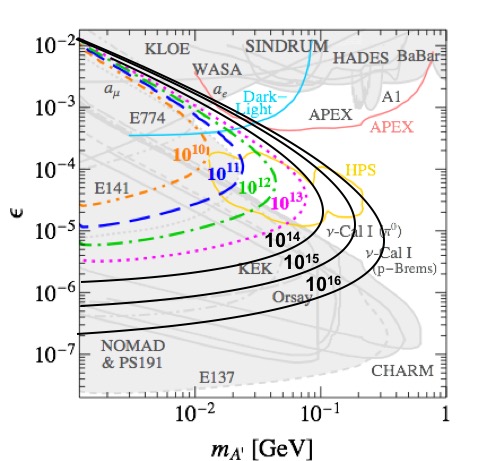}
\caption{Limits on dark photon production decaying to an $e^+e^-$ pair in terms of the mixing strength, $\epsilon$ and dark photon photon 
mass, $m_{A^\prime}$, from previous measurements (light grey shading).  The expected sensitivity for the NA64 experiment is shown for a 
range of electrons on target, $10^{10} - 10^{13}$.  Rough estimates are also shown for $10^{14} - 10^{16}$ electrons on target which could 
be provided to an NA64-like experiment by a future AWAKE accelerator scheme.}
\label{fig_sensitivity}
\end{figure}

The NA64 experiment is already making significant progress investigating new regions of phase space for dark photons and as shown 
in Fig.~\ref{fig_sensitivity} will cover much new ground in the $\epsilon-m_{A^\prime}$ plane.  Given the limitations of the number of 
electrons on target, the AWAKE acceleration scheme could make a real impact as the number of electrons is expected to be several 
orders of magnitude higher.  Assuming a bunch of $10^9$ electrons produced every 5\,s and a running period of 3\,months gives 
$10^{15}$ electrons on target and this along with a range of other values is shown in Fig.~\ref{fig_sensitivity}.  Further studies are ongoing 
and a higher number of electrons on target should be possible depending on the SPS injection scheme as well as the success of AWAKE 
in accelerating bunches of electrons.  The extra electrons on target provide extra reach into an unexplored region in the 
$\epsilon-m_{A^\prime}$ plane.

More detailed work is also ongoing to provide more robust estimates of the sensitivity as well as the ultimate number of electrons on target.  
A crucial difference to NA64's current set-up is that AWAKE produces short bunches of high charge.  This is one reason why the decay to 
$e^+e^-$ pairs is considered rather than the invisible mode.  Other channels such as the decay to a $\mu^+ \mu^-$ pair should also be 
considered.  Other aspects which need to be studied are optimising the experimental set up, re-assessing the backgrounds, particularly 
given the higher number of electrons on target, and considering the optimal beam energy.

\subsection{High-energy electron--proton/ion colliders}

A high-energy electron--proton/ion ($ep/eA$) facility could be the first application of plasma wakefield acceleration to particle colliders.  In such collisions, 
the electron generally emits a photon of virtuality $Q^2$ and strikes a parton carrying a fraction, $x$, of the proton's momentum.  The 
higher the $Q^2$, the smaller the probe and hence the more detailed structure can be seen; also low values of $x$ probe low momentum 
particles and hence the dynamic structure of quark and gluon radiation within the proton.  As such, $ep/eA$ collisions provide a detailed picture 
of the fundamental structure of matter and investigates the strong force of nature and its description embodied within quantum chromodynamics (QCD).  
Some of the open issues to be investigated in $ep/eA$ collisions are: when does this rich structure of gluon and quark radiation stop or "saturate" as 
it surely must otherwise cross sections would become infinite; in general, the nature of high-energy hadronic cross sections; and is there 
further substructure or are partons fundamental point-like objects.

Such an $ep$ collider 
would be more easily achieved than a high-luminosity, high-energy $e^+e^-$ collider for a number of reasons:  the proton bunches used in 
collisions are available in the CERN accelerator complex; an $ep$ machine could make use of positrons, as well as electrons, although this is 
not absolutely necessary; electron bunches of small transverse extent (nm-scale) are not necessary as the proton bunches are at the 
micron-scale; and certain processes and kinematic regions in $ep$ physics do not require high luminosity and so the physics case is still 
strong at low luminosity values.  Initial collider designs~\cite{nim:a740:173} considered generating electron bunches via the AWAKE scheme 
with electrons up to about 100\,GeV.  This has been formulated into the PEPIC (Plasma Electron--Proton/Ion Collider) project in which the 
SPS protons are used to drive wakefields and accelerate electrons to about 50\,GeV which then collide with LHC protons.  This would 
have essentially the same energy reach as the LHeC project~\cite{lhec}, but with luminosities several orders of magnitude lower.  As such, 
it would focus on studies of the structure of matter and QCD, in particular at low values of $x$ where the 
event rate is high.

Plasma simulations have shown the the LHC proton bunches can be used as a drive beam and accelerating gradients of just under 1\,GV/m 
are possible for long distances, leading to the possibility of accelerating electrons up to 6\,TeV in under 10\,km~\cite{pp:18:103101}.  Such 
values are unrealistic with conventional RF accelerators.  Given these promising results, a very high energy electron--proton (VHEeP)~\cite{epj:c76:463} 
collider has been proposed in which LHC bunches are used to drive wakefields and accelerate electrons to 3\,TeV in under 4\,km, which then 
collide with the counter-propagating proton (or ion) bunch, creating electron--proton collisions at centre-of-mass energies, $\sqrt{s}$, of over 9\,TeV.  The 
energies of the electrons could be varied, although the distance of 4\,km fits comfortably within the circumference of LHC ring, so although 
there maybe an upper energy limit, lower energies should be achievable.  Such centre-of-mass energies represent a factor of 30 increase compared 
to HERA which allows an extension to low $x$ and to high $Q^2$ of a factor of 1000.  The luminosity is 
currently estimated to be around 
$10^{28} - 10^{29}$\,cm$^{-2}$\,s$^{-1}$ which would lead to an integrated luminosity of 1\,pb$^{-1}$ per year.  Different schemes to improve this 
value are being considered such as squeezing the proton (and electron) bunches, multiple interaction points, etc.  However, even at these modest 
luminosities, such a high-energy electron--proton collider has a strong physics case.

\begin{figure}[!h]
\centering\includegraphics[width=12cm]{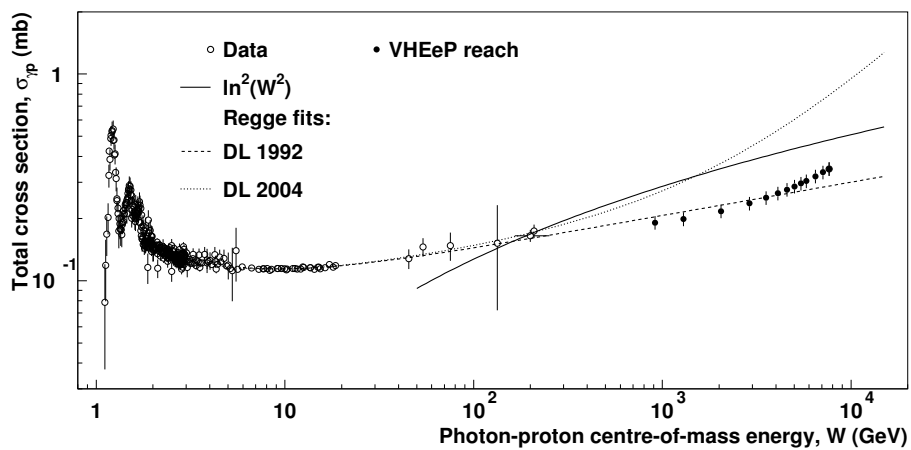}
\caption{Total $\gamma p$ cross section versus photon--proton centre-of-mass energy, $W$, shown for data compared to 
various models~\cite{pl:b296:227,Donnachie:2004pi,pr:123:1053}.  The data is taken from the PDG~\cite{pdg}, with 
references to the original papers given therein. The highest VHEeP data point is shown at $y=0.7$ (where $y = W^2/s$); the 
cross section at this point is assumed to be double the ZEUS value.  The other VHEeP points assume different values of 
$y$  down to 0.01 and are plotted on a straight line (linear in $W$) between the ZEUS and highest VHEeP point.  All VHEeP 
points have the same uncertainty as the ZEUS point: a systematic uncertainty of 7.5\% and a negligible statistical uncertainty.  
The ZEUS measurement is at $\sqrt{s} = 209$\,GeV and used a luminosity of 49\,nb$^{-1}$. This result has been updated 
from the original paper~\cite{epj:c76:463} with the addition of newly-calculated points for VHEeP.}
\label{fig_sigtot}
\end{figure}

The physics potential of VHEeP was discussed in the original publication~\cite{epj:c76:463} and discussed further at a dedicated 
workshop~\cite{vheep-ws} on the subject.  An example and updated result is shown in Fig.~\ref{fig_sigtot}, in which the total $\gamma p$ 
cross section is shown versus the photon--proton centre-of-mass energy, $W$.  This is a measurement which relies on only a modest 
luminosity and will be dominated by systematic uncertainties.  As can be seen from the expected VHEeP data, the measurement is extended 
to energies well beyond the current data, into a region where the dependency of the cross section is not known.  Some models are also shown 
and they clearly differ from each other at the high energies achievable at VHEeP.  These data could also be useful in 
understanding more about cosmic-ray physics as such collisions correspond to a 20\,PeV photon on a fixed target.

Several other possible measurements were presented in the original paper~\cite{epj:c76:463} in which the current theories of QCD are completely 
inadequate and where this new collider will significantly constrain our understanding of the fundamental structure of matter.  As for  
Fig.~\ref{fig_sigtot}, cross sections must at some point stop rising and cannot be arbitrarily large at high energies.  Exactly when a different behaviour 
will occur is not easy to predict, but there are strong indications that VHEeP energies will be sufficient to observe changes in several 
cross-section measurements such as the production of vector mesons and the inclusive $\gamma p$ and $\gamma^* p$ cross sections.  The 
theory of hadronic interactions at high energies and the nature of QCD and structure of matter was also extensively discussed at the dedicated workshop 
on the subject~\cite{vheep-ws}.

At the very highest $Q^2$ values, searches for high energy phenomenon beyond the Standard Model will be possible.  Higher luminosities 
will allow a comprehensive search to complement those at the LHC; however, even with modest luminosities, some specific processes can 
be investigated with higher sensitivity at VHEeP than at the LHC.  As an example, the production of leptoquarks, which would be produced 
on mass shell, is possible up to the kinematic limit of the centre-of-mass energy, i.e.\ 9\,TeV.  Other examples of physics that could be 
investigated at VHEeP were presented and discussed at the workshop\cite{vheep-ws}.  It was discussed how high-energy $ep$ 
collisions are sensitive to new descriptions and general theories of particle interactions~\cite{dvali} as well as having connections with 
black holes and gravity~\cite{erdmenger}.  

\section{Summary}
\label{sec:summary}

The AWAKE collaboration has during 2018 successfully demonstrated proton-driven plasma wakefield acceleration for the first time.  A 
programme of R\&D is being developed to be able to apply this scheme as a useable accelerator technology, hence the emphasis is 
the use of the CERN proton accelerators to drive wakefields and in general how this can fit into the CERN infrastructure.  Given this, 
realistic applications of particle physics experiments are being considered, most notably: the investigation of QED in strong fields; a 
beam-dump experiment to search for dark photons; and an electron--proton collider at the highest energies to investigate the fundamental 
structure of matter.  As well as being able to integrate these into the CERN accelerator complex and providing a testing ground for 
the accelerator technology, it is important that the experiments will also investigate new areas of particle physics .  Overall, this paper 
presents a path in which a new accelerator technology of ever-increasing performance, such as high energy 
or luminosity and better beam quality, can be applied to a new generation of particle experiments.

\enlargethispage{20pt}


\dataccess{The article has no additional data.}


\competing{The author declares that they have no competing interests.}

\funding{This work was supported by a Leverhulme Trust Research Project Grant RPG-2017-143.}

\ack{
The author also acknowledges the support of the Alexander von Humboldt Stiftung and DESY, Hamburg.  The author would like to thank 
E. Adli, A. Caldwell, P. Crivelli, S. Gninenko, E. Gschwendtner, A. Hartin, B. Heinemann, F. Keeble and P. Muggli for helpful discussions 
on the ideas contained here.  
Such new particle physics experiments are only conceivable because of the tremendous success of the AWAKE experiment and hence the 
collaboration deserves special mention.}




\begin{thebibliography}{9}

\bibitem{np:22:579}
Glashow SL. 1961. Partial symmetries of weak interactions.
\textit{Nucl. Phys.} \textbf{22}, 579.

\bibitem{prl:19:1264}
Weinberg S. 1967. A model of leptons.
\textit{Phys. Rev. Lett.} \textbf{19}, 1264.

\bibitem{salam-nobel}
Salam A. 1968. Weak and electromagnetic interactions.
In proceedings of the eighth Nobel symposium, \textit{Elementary Particle Physics: Relativistic Groups and Analyticity}.
Almquvist and Wiksell (Eds.), p.367.

\bibitem{np:b44:189}
't Hooft G, Veltman M. 1972. Regularization and renormalization of gauge fields.
\textit{Nucl. Phys.} \textbf{B~44}, 189.

\bibitem{prl:13:321}
Englert F, Brout R. 1964.  Broken symmetry and the mass of gauge vector mesons. 
\textit{Phys. Rev. Lett.} \textbf{13}, 321.

\bibitem{pl:12:132}
Higgs, PW. 1964. Broken symmetries, massless particles and gauge fields.
\textit{Phys. Lett.} \textbf{12}, 132.

\bibitem{prl:13:508}
Higgs, PW. 1964. Broken symmetries and the masses of gauge bosons.
\textit{Phys. Rev. Lett.} \textbf{13}, 508.

\bibitem{prl:13:585}
Guralnik GS, Hagen CR, Kibble TWB, 1964.  Global conservation laws and massless particles.
\textit{Phys. Rev. Lett.} \textbf{13}, 585.

\bibitem{pr:145:1156}
Higgs, PW. 1966. Spontaneous symmetry breakdown without massless bosons. 
\textit{Phys. Rev.} \textbf{145}, 1156.

\bibitem{pr:155:1554}
Kibble, TWB. 1967. Symmetry breaking in non-Abelian gauge theories. 
\textit{Phys. Rev.} \textbf{155}, 1554.

\bibitem{pl:b716:1}
ATLAS Coll., Aad G, et al. 2012. Observation of a new particle in the search for the Standard Model Higgs boson 
with the ATLAS detector at the LHC.
\textit{Phys. Lett.} \textbf{B~716}, 1.

\bibitem{pl:b716:30}
CMS Coll., Chatrchyan S, et al. 2012. Observation of a new boson at a mass of 125 GeV with the CMS experiment 
at the LHC .
\textit{Phys. Lett.} \textbf{B~716}, 30.

\bibitem{lhc}
Evans L, Bryant P. 2008. LHC machine.
\textit{J. Instrum.} \textbf{3}, S08001.

\bibitem{ilc}
The ILC Technical Design Report, \\{\tt http://www.linearcollider.org/ILC/Publications/Technical-Design-Report}

\bibitem{clic}
Burrows P, et al. (Eds.) 2016. Updated baseline for a staged Compact Linear Collider. 
CERN-2016-004, CERN.

\bibitem{Tajima:1979bn}
Tajima T, Dawson JM. 1979. Laser electron accelerator.
\textit{Phys. Rev. Lett.} \textbf{43}, 267.

\bibitem{Chen:1984up}
Chen P, Dawson JM, Huff RW, Katsouleas TC. 1985. Acceleration of electrons by the interaction of a bunched electron 
beam with a plasma.
\textit{Phys. Rev. Lett.} \textbf{54}, 693.

\bibitem{pop-14-055501}
Joshi C. 2007. The development of laser- and beam-driven plasma accelerators as an experimental field.
\textit{Phys. Plasmas} \textbf{14}, 055501.

\bibitem{RevModPhys.81.1229}
Esarey E, Schroeder CB, Leemans WP. 2009.  Physics of laser-driven plasma-based electron accelerators.
\textit{Rev. Mod. Phys.} \textbf{81}, 1229.

\bibitem{doi:10.1142/S1793626816300036}
Hogan MJ. 2017.  Electron and positron beam-driven plasma acceleration.
\textit{Rev. Acc. Sci. Tech.} \textbf{9}, 63.

\bibitem{Modena:1995dgw}
Modena A, et al. 1995. Electron acceleration from the breaking of relativistic plasma waves.
\textit{Nature} \textbf{377}, 606.

\bibitem{Mangles:2004ta}
Mangles SPD, et al. 2004. Monoenergetic beams of relativistic electrons from intense laser--plasma interactions.
\textit{Nature} \textbf{431}, 535.

\bibitem{Geddes:2004tb}
Geddes CGR, et al. 2004.  High-quality electron beams from a laser wakefield accelerator using plasma-channel guiding.
\textit{Nature} \textbf{431}, 538.

\bibitem{Faure:2004tc}
Faure J, et al. 2004.  A laser--plasma accelerator producing monoenergetic electron beams.
\textit{Nature} \textbf{431}, 541.

\bibitem{Blumenfeld:2007ph}
Blumenfeld I, et al. 2007.  Energy doubling of 42 GeV electrons in a metre-scale plasma wakefield accelerator.
\textit{Nature} \textbf{445}, 741.

\bibitem{Litos:2014}
Litos M, et al. 2014.  High-efficiency acceleration of an electron beam in a plasma wakefield accelerator.
\textit{Nature} \textbf{515}, 92.

\bibitem{Caldwell:2008ak}
Caldwell A, Lotov K, Pukhov A, Simon F. 2009. Proton-driven plasma-wakefield acceleration.
\textit{Nature Phys.} \textbf{5}, 363.

\bibitem{ppcf:56:084013} 
AWAKE Coll., Assmann R, et al. 2014. Proton-driven plasma wakefield acceleration: a path to the future of 
high-energy particle physics. 
\textit{Plasma Phys. Contr. Fusion} \textbf{56}, 084013.

\bibitem{nim:a829:3}
AWAKE Coll., Caldwell A, et al. 2016. Path to AWAKE: Evolution of the concept. 
\textit{Nucl. Instrum. Meth.} \textbf{A~829}, 3. 

\bibitem{nim:a829:76}
AWAKE Coll., Gschwendtner E, et al. 2016. AWAKE, The Advanced Proton Driven Plasma Wakefield 
Acceleration Experiment at CERN. \textit{Nucl. Instrum. Meth.} \textbf{A~829}, 76.

\bibitem{ppcf:60:014046}
AWAKE Coll., Muggli P, et al. 2018. AWAKE readiness for the study of the seeded self-modulation of a 
400\,GeV proton bunch. 
\textit{Plasma Phys. Control. Fusion} \textbf{60}, 014046. 

\bibitem{arxiv:1809.01191}
AWAKE Coll., Turner M, et al. 2018. Experimental observation of plasma wakefield growth driven by the 
seeded self-modulation of a proton bunch.
arXiv:1809.01191.

\bibitem{arxiv:1809.04478}
AWAKE Coll., Adli E, et al. 2018. Experimental observation of proton bunch modulation in a plasma, at varying 
plasma densities.
arXiv:1809.04478.

\bibitem{nature:2018}
AWAKE Coll., Adli E, et al. 2018. Acceleration of electrons in the plasma wakefield of a proton bunch. 
\textit{Nature} \textbf{561}, 363.

\bibitem{adli:ipac16}
Adli E. 2016. Towards AWAKE applications: electron beam acceleration in a proton driven plasma wake.
\textit{Proceedings of IPAC2016}, Busan, Korea, p.2557.




\bibitem{dis:book}
See for example: Devenish R, Cooper-Sarkar A. 2004 \textit{Deep Inelastic Scattering}.  Oxford University Press, Oxford, UK.

\bibitem{hera}
HERA - A Proposal for a Large Electron Proton Colliding Beam Facility at DESY. 1981. DESY-HERA-81-10.

\bibitem{mpl:a24:1087}
See for example: Bass SD. 2009. The proton spin puzzle: Where are we today ?
\textit{Mod. Phys. Lett.} \textbf{A~24}, 1087.

\bibitem{ijmp:a33:1830011}
Hartin A. 2018. Strong field QED in lepton colliders and electron/laser interactions.
\textit{Int. J. Mod. Phys.} \textbf{A~33}, 1830011.

\bibitem{pr:d60:092004}
E144 Coll., Bamber C, et al. 1999.  Studies of nonlinear QED in collisions of 46.6-GeV electrons with intense laser pulses. 
\textit{Phys. Rev.} \textbf{D~60}, 092004.

\bibitem{sqed-ws}
See talks at workshop on "Probing strong-field QED in electron--photon interactions". 2018. DESY, {\tt https://indico.desy.de/indico/event/19493/}

\bibitem{JETP:56:502}
Okun LB. 1982. Limits of electrodynamics: Paraphotons?
\textit{Sov. Phys. JETP} \textbf{56}, 502 [\textit{Zh. Eksp. Teor. Fiz.} \textbf{83}, 892].

\bibitem{pl:b136:279}
Galison P, Manohar A. 1984. Two $Z$s or not two $Z$s.
\textit{Phys. Lett.} \textbf{B~136}, 279.

\bibitem{pl:b166:196}
Holdom B. 1986. Two U(1)s and epsilon charge shifts.
\textit{Phys. Lett.} \textbf{B~166}, 196.

\bibitem{arnps:60:405}
Jaeckel J, Ringwald A. 2010. The low-energy frontier of particle physics.
\textit{Ann. Rev. Nucl. Part. Sci.} \textbf{60}, 405.

\bibitem{us-cosmic-visions}
Battaglieri M, et al. 2017. US Cosmic Visions: New Ideas in Dark Matter 2017: Community Report.
arXiv:1707.04591.

\bibitem{pr:d89:075008}
Gninenko S. 2014. Search for MeV dark photons in a light-shining-through-walls experiment at CERN.
\textit{Phys. Rev.} \textbf{D~89}, 075008.

\bibitem{arxiv:1312.3309}
Andreas S, et al. 2013.  Proposal for an experiment to search for light dark matter at the SPS.
arXiv:1312.3309.

\bibitem{prl:118:011802}
NA64 Coll., Banerjee D, et al. 2017. Search for invisible decays of sub-GeV dark photons in missing-energy events 
at the CERN SPS.
\textit{Phys. Rev. Lett.} \textbf{118}, 011802.

\bibitem{nim:a740:173}
Xia G, et al. 2014. Collider design issues based on proton-driven plasma wakefield acceleration.
\textit{Nucl. Instrum. Meth.} \textbf{A~740}, 173. 

\bibitem{lhec}
LHeC Study Group, Abelleira Fernandez JL, et al. 2012. A Large Hadron Electron Collider at CERN.
\textit{J. Phys.} \textbf{G~39}, 075001.

\bibitem{pp:18:103101}
Caldwell A, Lotov K. 2011. Plasma wakefield acceleration with a modulated proton bunch.
\textit{Phys. Plasmas} \textbf{18}, 103101.

\bibitem{epj:c76:463} 
Caldwell A, Wing M. 2016. VHEeP: A very high energy electron--proton collider.  
\textit{Eur. Phys. J.} \textbf{C~76}, 463.

\bibitem{vheep-ws}
See talks at workshop on "Prospects for a very high energy $eP$ and $eA$ collider and Leo Stodolsky symposium". 2017. 
Max Planck Institute for Physics, Munich, {\tt https://indico.mpp.mpg.de/event/5222/overview}

\bibitem{pl:b296:227}
Donnachie A, Landshoff PV. 1992. Total cross-sections. \textit{Phys. Lett.} \textbf{B~296}, 227.

\bibitem{Donnachie:2004pi}
Donnachie A, Landshoff PV. 2004. Does the hard pomeron obey Regge factorization. \textit{Phys. Lett.} \textbf{B~595}, 393. 

\bibitem{pr:123:1053}
Froissart M. 1961. Asymptotic behavior and subtractions in the Mandelstam representation. \textit{Phys. Rev.} \textbf{123}, 1053.

\bibitem{pdg}
Particle Data Group, Yao W-M, et al. 2006. The review of particle physics. \textit{J. Phys.} \textbf{G~33}, 1; data from 
{\tt http://pdg.lbl.gov/2006/hadronic-xsections/hadron.html}

\bibitem{dvali}
Dvali, G. 2017. High energy cross sections and classicalization. Talk presented at~\cite{vheep-ws}.

\bibitem{erdmenger}
Erdmenger, J. 2017. Applications of AdS/CFT to very low-$x$ physics. Talk presented at~\cite{vheep-ws}.

\end{thebibliography}
\end{document}